\documentclass[sigconf]{acmart}
\AtBeginDocument{%
  }

\setcopyright{acmlicensed}
\acmDOI{XXXXXXX.XXXXXXX}
\acmISBN{978-1-4503-XXXX-X/18/06}

\usepackage{booktabs}
\usepackage{graphicx}
\usepackage{titlesec}
\usepackage{svg}
\titlespacing*{\section}{0pt}{4pt}{4pt} 
\titlespacing*{\subsection}{0pt}{3pt}{3pt} 
\usepackage{amsmath}
\usepackage{booktabs}
\usepackage{multirow} 
\usepackage{balance}
\usepackage{xcolor}
\usepackage{colortbl}  
\titlespacing*{\section}{0pt}{*1}{*1} 
\titlespacing*{\subsection}{0pt}{*1}{*1} 
\copyrightyear{2025}
\acmYear{2025}
\setcopyright{cc}
\setcctype{by}
\acmConference[SIGIR '25]{Proceedings of the 48th International ACM SIGIR Conference on Research and Development in Information Retrieval}{July 13--18, 2025}{Padua, Italy}
\acmBooktitle{Proceedings of the 48th International ACM SIGIR Conference on Research and Development in Information Retrieval (SIGIR '25), July 13--18, 2025, Padua, Italy}\acmDOI{10.1145/3726302.3730223}
\acmISBN{979-8-4007-1592-1/2025/07}

\begin{document}

\title{LLM-Driven Usefulness Labeling for IR Evaluation}

%
\author{Mouly Dewan}
\affiliation{%
  \institution{University of Washington}
  \city{Seattle, WA}
  \country{United States}}
\email{mdewan@uw.edu}

\author{Jiqun Liu}
\affiliation{%
  \institution{The University of Oklahoma}
  \city{Norman, OK}
  \country{United States}}
\email{jiqunliu@ou.edu}

\author{Chirag Shah}
\affiliation{%
  \institution{University of Washington}
  \city{Seattle, WA}
  \country{United States}}
\email{chirags@uw.edu}

\begin{abstract}
In the information retrieval (IR) domain, evaluation plays a crucial role in optimizing search experiences and supporting diverse user intents. In the recent LLM era, research has been conducted to automate document relevance labels. These labels have traditionally been assigned by crowd-sourced workers, a process that is both time consuming and costly. This study focuses on LLM-generated \textit{usefulness} labels, a crucial evaluation metric that considers the user's search intents and task objectives, an aspect where relevance falls short. Our experiment utilizes task-level, query-level, and document-level features along with user search behavior signals, which are essential in defining the usefulness of a document. Our research finds that (i) pre-trained LLMs can generate moderate usefulness labels by understanding the comprehensive search task session, and (ii) pre-trained LLMs perform better judgment in short search sessions when provided with search session contexts. Furthermore, we investigate whether LLMs can capture the unique divergence between relevance and usefulness, along with conducting an ablation study to identify the most critical metrics for accurate usefulness label generation. In conclusion, this work explores LLM-generated usefulness labels by evaluating critical metrics and optimizing for practicality in real-world settings.
\end{abstract}



\maketitle

\section{Introduction}
Assessing relevance is not enough for many IR scenarios, as in many cases it cannot fully capture the extent to which a document is actually \textit{useful} for addressing the motivating tasks. Although many scholars have advocated this for decades \cite{cole2009usefulness, liu_toward_2022}, mainstream IR has remained focused on the seemingly objective assessment of relevance, following the traditional Cranfield evaluation paradigm \cite{cormack1998efficient}. Relevance is often used as a proxy for other constructs due to its strong correlation with user centric evaluation labels like usefulness and satisfaction~\cite{mao_when_2016}. However, usefulness is tied to both task context and users' context dependent preferences, making it a richer but more challenging construct to capture in real time. For web search evaluations, relevance based metrics alone cannot capture the user’s search goal and satisfaction, especially during complex task driven sessions \cite{belkin2008relevance, liu2021deconstructing}. Therefore, incorporating usefulness as an evaluation metric provides a more holistic assessment, capturing not only document relevance, but also user search intent and satisfaction throughout sessions~\cite{mitsui2017predicting, liu2020identifying}. 

To mitigate time and cost constraints of traditional evaluation challenges \cite{thomas_large_2024}, IR researchers have been exploring the possibility and techniques of automated search evaluation with LLMs \cite{faggioli_perspectives_2023, chen_ai_2024, alaofi2024llms}. LLMs for automated evaluation (LLM4eval) have recently attracted growing attention in IR community because of their flexibility and increasing task-solving capabilities \cite{rahmani_llm4eval_2024}. In LLM automated evaluations, researchers have primarily focused on relevance labels, while user centric usefulness metrics remain largely unexplored. As usefulness is a measure that directly reflects a \textit{document's contribution to achieving user goals}, in this paper we particularly focus on usefulness as the key metric for evaluating web search along with relevance. To achieve this, we generate usefulness labels using LLMs and employ search session, task-level, query-level and document-level user metrics, aiming to better capture how users perceive the quality of search systems. We also seek to explore whether LLMs can capture the diverging relationship between relevance and usefulness. In some cases, a retrieved document may be relevant to the query but not necessarily useful to the user, or it may be irrelevant to the query yet highly valuable for completing the task. Being able to identify and address such situations can open up new possibilities for enhancing IR applications, especially in complex, open ended tasks. 

Existing ML-based approaches for predicting usefulness often overlook the broader task and user contexts. This is the place where LLMs can step in with their pre-trained existing knowledge to better understand user, task, and query context through appropriate prompting. The purpose of generating usefulness labels is to gauge how well and accurately these LLMs can perform in automated usefulness judgment when given all relevant features (both implict and explicit) \cite{chen2017meta, zhang2018well, liu2018satisfaction, fox2005evaluating}. Specifically, our research seeks to answer the following three research questions (RQs):

{\bfseries RQ1}: To what extent can large language models (LLMs) accurately label document \textit{usefulness} in search sessions?

{\bfseries RQ2}: To what extent LLMs understand the comprehensive search goal of a user and navigate the divergence between relevance and usefulness of the retrieved documents?

{\bfseries RQ3}: What are the most effective search metrics in improving LLMs' document usefulness judgments?

\section{Related Works}

In LLM enabled search engine evaluation, most of the research has solely focused on automating relevance labels \cite{abbasiantaeb_can_2024, faggioli_perspectives_2023, rahmani2024llmjudge, upadhyay_large-scale_2024}, where else user centric usefulness metrics have not been explored yet. Faggioli et al. \cite{faggioli_perspectives_2023} were among the first to examine how LLMs can be used for relevance judgment, offering unique insights into human-LLM collaboration. Their work explored direct prompting strategies, highlighting both the potential and limitations of using LLMs to generate relevance labels. Another prominent work in LLM automated relevance assessment by Thomas et al. \cite{thomas_large_2024} asserted that LLMs can accurately predict searcher’s preferences and perform relevance labeling as accurately as human labelers. Studies conclude that LLMs are more positive in assessing relevance and thus should be used to assess all documents, not just relevance gaps \cite{abbasiantaeb_can_2024} and, to some extent LLMs are studied for utility judgment using retrieval-augmented generation (RAG) \cite{zhang_are_2024, salemi2024evaluating}. To address some limitations of Thomas et al. \cite{thomas_large_2024}, Upadhyay et al. \cite{upadhyay_large-scale_2024} conducted large scale relevance assessment on DL TREC tracks using UMBRELLA \cite{upadhyay_umbrela_2024}, an LLM based relevance assessor for academic TREC-style evaluation. Most LLM automated IR evaluations have focused solely on relevance labeling, but relevance alone is insufficient, as usefulness directly reflects user goal achievement \cite{belkin2008relevance, cole2009usefulness}. In this work, we explore usefulness as a key metric for web search evaluation alongside relevance. We generate usefulness labels using LLMs, incorporating implicit user search behavior signals like CTR and dwell time \cite{chuklin2013click, chapelle2009expected, jiang2015understanding}. Additionally, we examine whether LLMs can capture contextual relationships in search sessions by integrating relevance and satisfaction labels to assess search system quality while exploring the contrast between relevance and usefulness.

\section{Methodology} 
Our main goal is to evaluate the extent to which LLMs can generate usefulness labels when given user search behavior signals along with task-level, query-level and document-level information. In real-world scenarios, when evaluation is conducted by crowd-sourced workers they perform individual search sessions for each task and assign usefulness labels for specific query and task within the context of the entire search session. In our experiment, we follow the same methodology where LLMs are provided vital information about each user’s search session and it generates a usefulness label for each task and for all the clicked documents. 

We conduct the experiments using our zero-shot prompt\footnotemark[1]\footnotetext[1]{https://github.com/moulydewan/LLMUsefulnessLabels} that has been created taking inspiration from the structure of the  widely used DNA prompt template by Thomas et al. \cite{thomas_large_2024} that uses LLMs to generate relevance labels. We conduct experiments on 6 different types of pre-trained LLMs such as GPT-4omini, GPT-3.5turbo \cite{achiam2023gpt}, Llama 3.1 8B-Instruct, Llama 3.2 3B-Instruct, Llama 3.3 70B-Instruct \cite{dubey2024llama} and DeepSeek-R1-Distill-Qwen-14B \cite{liu2024deepseek} to evaluate the ability to generate usefulness labels. Additionally, we explore LLM's ability to process the divergence cases of \textit{relevance-usefulness} to capture the nuances of search session contexts. Finally, our research includes an ablation study of various search behavior metrics to identify the most effective ones for input into an LLM for optimizing accuracy while considering real-world constraints, such as token limitations, model sizes and associated costs.


\begin{table}[h]
    \caption{Statistics of Behavior Logs}
    \label{tab:behavior_logs}
    \centering
    \resizebox{\linewidth}{!}{ 
    \begin{tabular}{lccccc}
        \toprule
        \textbf{Dataset} & \textbf{\#Tasks} & \textbf{\#Participants} & \textbf{\#Sessions} & \textbf{\#Queries} & \textbf{\#Clicks} \\ \midrule
        THUIR-KDD'19 & 9  & 50    & 447   & 735   & 1,431  \\ 
        QRef      & --- & 42   & 2,024 & 4,809 & 7,126  \\ 
        \bottomrule
    \end{tabular}
    }
\end{table}

\begin{table*}[t]
\caption{Comparison of LLM performances on THUIR-KDD'19 and QRef datasets using Spearman's Rank ($\rho$).}
\label{tab:model_performance}
\centering
\small
\setlength{\tabcolsep}{4.3pt} 
\begin{tabular}{lccc|ccc|ccc|ccc}
\toprule
\multirow{2}{*}{\textbf{Model}} & \multicolumn{6}{c|}{\textbf{THUIR-KDD'19 Dataset}} & \multicolumn{6}{c}{\textbf{QRef Dataset}} \\
 & \multicolumn{3}{c}{\textbf{Baseline}} & \multicolumn{3}{c|}{\textbf{Session}} & \multicolumn{3}{c}{\textbf{Baseline}} & \multicolumn{3}{c}{\textbf{Session}} \\
 & \textbf{Overall} & \textbf{Task} & \textbf{Query} & \textbf{Overall} & \textbf{Task} & \textbf{Query} & \textbf{Overall} & \textbf{Session} & \textbf{Query} & \textbf{Overall} & \textbf{Session} & \textbf{Query} \\
\midrule
GPT-4omini    & 0.35 & 0.30 & 0.28 & \textbf{0.36*} & 0.30 & 0.25 & 0.29 & 0.35 & \textbf{0.44+} & 0.37 & 0.41 & \textbf{0.48*} \\
GPT-3.5turbo  & \textbf{0.37+}  & 0.31 & 0.26 & 0.33 & 0.28 & 0.26 & 0.21 & 0.34 & 0.42 & 0.37 & 0.45 & \textbf{0.47*} \\
Llama 3.1 8B-Instruct  & 0.31 & 0.26 & 0.20 & 0.31 & 0.27 & 0.27 & 0.25 & 0.29 & 0.39 & 0.27 & 0.32 & 0.39 \\
Llama 3.2 3B-Instruct  & 0.19 & 0.12 & 0.06 & 0.22 & 0.17 & 0.21 & 0.13 & 0.23 & 0.32 & 0.30 & 0.37 & 0.46 \\
Llama 3.3 70B-Instruct & \textbf{0.38+} & 0.34 & 0.30 & \textbf{0.37*} & 0.32 & 0.28 & 0.33 & 0.38 & \textbf{0.44+} & 0.37 & 0.41 & 0.42 \\
DeepSeek-R1-Distill-Qwen-14B & 0.33 & 0.27 & 0.22 & - & - & - & 0.15 & 0.13 & 0.11 & - & - & - \\
\bottomrule
\end{tabular}
\end{table*}

\subsection{Datasets}
For this experiment, we used two datasets: THUIR KDD’19 \cite{mao2016does} and TianGong-QRef \cite{chen2021towards}, with behavioral log statistics shown in Table~\ref{tab:behavior_logs}. Both datasets include query satisfaction, session satisfaction, and usefulness labels while only THUIR-KDD’19 provides relevance labels from external assessors. They capture user search interactions (CLICK, SCROLL, HOVER, MOVE) and session timestamps. From these, we extracted CTR, URL dwell time, query dwell time, and task dwell time per user. From the statistics, we can see that THUIR-KDD’19 contains longer search sessions with 9 tasks, whereas QRef features shorter task-free sessions. For this reason, we perform task-level and query-level analysis for THUIR-KDD'19 and for QRef we perform session-level and query-level analysis (further shown in Table~\ref{tab:model_performance}). The variety in these datasets provided a diverse foundation for our study, allowing us to explore real-world user-centric search behaviors across both structured task-driven sessions and more open ended task-free search interactions.

\subsection{Baseline Data and Session Data} After curating all the additional metrics, we create our LLM dataset by treating each data point as an individual batch regardless of sessions and call them baseline data. Each data point represents a single user click on a web document with the associated metrics irrespective of user sessions. All corresponding queries, along with the metrics for each clicked document within a user’s task, are grouped into a single batch, which we refer to as session data. The reason for aggregating all these information together is that in practical scenario, the overall session history matters when labeling usefulness for each clicked documents. For a particular task, users typically either find the required information from the first query and the corresponding clicked documents, successfully completing their search, or they need to reformulate their query to refine the search results and then locate the desired information. If a user finds document X useful, a similar relevant document Y seen later may feel redundant and not useful. Yet without session context, the LLM might still rate both X and Y as equally useful. For this reason, to preserve the contextual information of a search session for a task, we formulate one batch as one task session for one user and provide it to the LLMs as a part of session data. Formulating each search session for each user in Table~\ref{tab:behavior_logs}, we get around 447 sessions from THUIR-KDD'19 dataset and 2,024 sessions from QRef dataset.

\section{Experimental Setup}
We conduct our primary experiment on two datasets: (1) baseline data without session context and (2) session data with previous search context. Both include document-level metrics (e.g., URL, title, summary, SERP snippets, rank) and behavioral metrics (e.g., CTR, URL-query-task dwell time), along with relevance and satisfaction labels for a comprehensive view of user interactions. Using our zero-shot prompting, we asked LLMs to label clicked document usefulness: 3 = Very Useful, 2 = Fairly Useful, 1 = Somewhat Useful, 0 = Not Useful at All. We evaluate OpenAI’s GPT-3.5 Turbo, GPT-4omini, and Meta’s latest open-source LLMs (Llama 3.3 70B-Instruct, Llama 3.2 3B-Instruct, Llama 3.1 8B-Instruct) \cite{dubey2024llama} for usefulness judgment. For LLM coverage, we included the reasoning model DeepSeek-R1-Distill-Qwen-14B along with other LLMs. Experiments were conducted via AWS Bedrock for Llama models and OpenAI APIs for GPT models, using a temperature of 0 and top\_p of 1. DeepSeek-R1-Distill-Qwen-14B was hosted locally via Ollama with the same temperature parameter.

\section{Results}
Our experiments address the three RQs for baseline and session data and evaluate LLM-generated usefulness at the overall level across all user interactions, while task and query scores are averaged separately to capture different granularity of ranking quality.

\subsection{RQ1: Baseline and Session Analysis}
\begin{figure}[t]
  \centering
  \includegraphics[width=\linewidth]{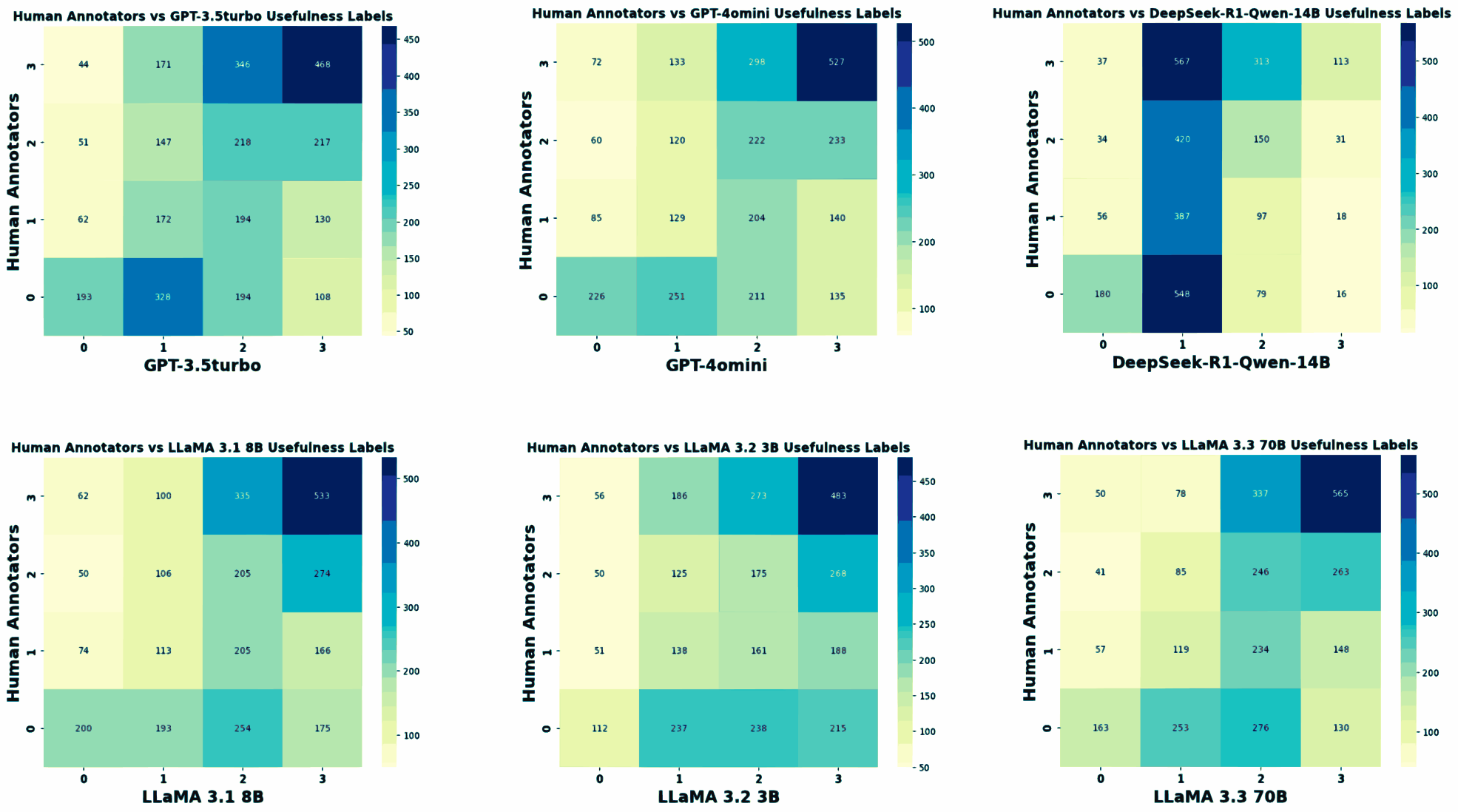} 
  \caption{Human annotators vs. LLM usefulness labels.}
  \label{fig:correlation}
  \Description{A comparison between human annotators and LLM-based usefulness labels.}
\end{figure}

We prompt the LLM models with our prompt and generate usefulness labels for each of the clicked documents. Both datasets already contain the usefulness labels annotated by humans that we use as our ground truth. To address our {\bfseries RQ1}, we evaluate how well the LLM-generated usefulness performs with respect to our ground truth. As the usefulness labels are ordinal, starting from 0 (low) to 3 (high), we conduct Spearman’s rank correlation (\(\rho\)) to measure association between ranked variables. The detailed analysis in Table~\ref{tab:model_performance} shows a moderate positive correlation between human and LLM-generated usefulness labels. The best performing models in both datasets are denoted with \textbf{+} for baseline and \textbf{*} for session. Considering {\bfseries Baseline} analysis, for THUIR-KDD'19 dataset for overall, Llama 3.3 70B-Instruct has the highest correlation (\(\rho\) = 0.38), followed by GPT-3.5turbo (\(\rho\) = 0.37). The weakest correlation is observed in Llama 3.2 3B-Instruct (\(\rho\) = 0.19), which is likely due to its relatively small parameter size (also shown in Figure~\ref{fig:correlation}). For the QRef dataset, the overall trend is similar with Llama 3.3 70B-Instruct performing moderately well (\(\rho\) = 0.33). Task-level and query-level findings for THUIR-KDD’19 align closely with overall trends, reinforcing consistency in our findings across all LLM models. In QRef’s baseline dataset for query level, both GPT-4omini (\(\rho\) = 0.44) and Llama 3.3 70B-Instruct (\(\rho\) = 0.44) perform the best. For {\bfseries Session} analysis, performance gradually improves compared to baseline, indicating that LLMs benefit from session-context search data. In QRef, improvement is observed in session compared to baseline in overall, task-level and query-level session data. This suggests that LLMs perform better in shorter (QRef) rather than longer (THUIR-KDD'19) search sessions. Notably, in QRef session data at the query level (\(\rho\) = 0.48 and 0.47), GPT models perform the best among all. This further supports the observation that session-level context enhances LLM performance, particularly in query-level analysis. Despite DeepSeek’s proficient reasoning ability, it has struggled in this task. During usefulness label generation, even after direct prompting instead of providing direct usefulness labels it often generated reasoning-based responses, requiring additional filtering and extraction of the labels from the generated responses. From our experiments, Llama 3.3 70B-Instruct remains the best performing model across different levels of analysis. Its 70B parameter count likely contributes to its superior ability to process and interpret nuanced contextual information. GPT-4omini follows next, performing well in both baseline and session contexts. Smaller models like Llama 3.2 3B-Instruct struggle to generate usefulness labels but improve performance with session data particularly in QRef's session data at both session (\(\rho\) = 0.37) and query level (\(\rho\) = 0.46).

\begin{table}[t]
    \caption{Comparison of LLM performances across 4 cases in baseline (B) and session (S) data.}
    \label{tab:model_performance1}
    \centering
    \resizebox{\columnwidth}{!}{ 
    \begin{tabular}{lcccccccc}
        \toprule
        \textbf{Model} & \multicolumn{2}{c}{\textbf{HR-HU}} & \multicolumn{2}{c}{\textbf{HR-LU}} & \multicolumn{2}{c}{\textbf{LR-HU}} & \multicolumn{2}{c}{\textbf{LR-LU}} \\
        \cmidrule(lr){2-3} \cmidrule(lr){4-5} \cmidrule(lr){6-7} \cmidrule(lr){8-9}
        & \textbf{B} & \textbf{S} & \textbf{B} & \textbf{S} & \textbf{B} & \textbf{S} & \textbf{B} & \textbf{S} \\
        \midrule
        GPT-3.5turbo & 0.10 & 0.12 & 0.02 & 0.09 & 0.17 & 0.04 & 0.07 & 0.05 \\
        GPT-4omini & 0.12 & 0.13 & -0.13 & 0.11 & 0.10 & 0.05 & 0.05 & 0.13 \\
        LLaMA 3.1 8B-Instruct & 0.09 & 0.11 & 0.05 & 0.03 & 0.12 & 0.04 & 0.04 & 0.11 \\
        LLaMA 3.2 3B-Instruct & 0.06 & 0.06 & 0.04 & 0.05 & 0.19 & 0.05 & 0.03 & 0.03 \\
        LLaMA 3.3 70B-Instruct & 0.16 & 0.14 & 0.13 & 0.08 & -0.02 & 0.06 & 0.11 & 0.11 \\
        DeepSeek-R1-Distill-Qwen-14B & 0.13 & - & 0.12 & - & 0.06 & - & 0.12 & - \\
        \bottomrule
    \end{tabular}
    }
\end{table}

\subsection{RQ2: Relevance vs Usefulness Divergence}
We analyze the correlation between relevance and usefulness across four cases: HR-HU (High Relevance, High Usefulness), HR-LU (High Relevance, Low Usefulness), LR-HU (Low Relevance, High Usefulness) and LR-LU (Low Relevance, Low Usefulness). As only THUIR-KDD'19 dataset has relevance labels we choose this dataset to conduct this part of experiment. We divide the dataset into four cases of usefulness-relevance combinations and tested the performance of LLMs. Both the relevance and usefulness labels are labeled from 0 to 3 where we consider 0 \& 1 as Low and 3 \& 4 as High. From our initial observation, we report the Spearman rank correlation (\(\rho\)) of the baseline and session data for 5 selected LLMs in Table~\ref{tab:model_performance1}. To address our {\bfseries RQ2}, we find that LLMs face challenges in capturing the contrasting relationship between relevance and usefulness in both baseline and session analysis. According to Spearman's rank correlation (\(\rho\)), all five LLM models are showing weak positive correlation and few of them have weak negative correlation. For the diverging cases HR-LU and LR-HU, the models face challenges in generating accurate usefulness labels. This may be due to the nature of Spearman's rank correlation, which relies on the overall rank order of data points. When the dataset is divided into four cases, the reduced range of rank variation and smaller sample sizes can weaken the correlation, making it harder to capture the relationship between human and LLM-generated usefulness labels within each subgroup. Furthermore, it shows that LLMs generalize measures such as high relevance will lead to high usefulness which is not the case always in real settings. The DNA structured prompt is solely not enough to capture the divergent cases, as such cases should be explicitly mentioned in the prompts for better clarification if expanded for future research.

\begin{table}[t]
\caption{Ablation study of features in GPT-4omini and LaMA 3.3 70B-Instruct with session data.}
\label{tab:ablation}
\centering
\begin{tabular}{lcc}
\toprule
\textbf{Features} & \textbf{GPT-4omini} & \textbf{LaMA 3.3 70B-Instruct} \\ 
\midrule
$U_{R+S+U}$ & \textbf{0.36} & \textbf{0.37} \\
$U_{R+S}$   & 0.35 & 0.35 \\
$U_{R+U}$   & 0.34 & 0.33 \\
$U_{S+U}$   & 0.34 & 0.31 \\
$U_R$       & 0.32 & 0.33 \\
$U_S$       & 0.27 & 0.29 \\
$U_U$       & 0.27 & 0.28 \\
\bottomrule
\end{tabular}
\end{table}

\subsection{RQ3: Ablation Study} 
To address our {\bfseries RQ3}, we conduct an ablation study of metrics provided to the LLM. We categorize the features into five levels: query features (Q), document features (D), relevance metrics (R), satisfaction metrics (S), and user behavior features (U) as shown in Table~\ref{tab:ablation}. For the ablation study, we keep Q and D constant while changing R, S and U. This approach aims to identify the most critical features required for LLMs to effectively generate usefulness labels. In real-world applications using all possible inputs in the LLM prompt is token-intensive and cost-prohibitive that's why it is important to optimize feature selection. For conducting the ablation study, we chose Llama 3.3 70B-Instruct and GPT-4omini on the THUIR-KDD'19 session data as these two consistently outperformed others in usefulness judgment. From the ablation study shown in Table~\ref{tab:ablation}, observing the Spearman's correlation (\(\rho\)) models seem to perform the best when all three features such as relevance (R), satisfaction (S) and user features (U) are provided along with query (Q) and document features (D) with a moderate positive correlation (\(\rho\) = 0.36 and 0.37). The next is followed by relevance (R) and satisfaction features (S) that also has a moderate positive correlation (\(\rho\) = 0.35). This shows that LLMs can infer usefulness if both relevance and satisfaction levels are present supporting prior research on usefulness \cite{mao2016does}. Notably, when satisfaction (S) and user features (U) are used individually it shows the lowest correlation (\(\rho\) = 0.27 and 0.28) indicating that both features are associated with usefulness should be considered for improved performance. The findings of the ablation study highlights relevance and satisfaction as key in determining usefulness and both combined can improve LLM's ability to capture nuanced contextual information. However, when features are provided independently, LLM struggles to effectively evaluate usefulness making it challenging to perform an accurate usefulness evaluation. This highlights the importance of combining features and focusing on relevance to enable the LLM to interpret and evaluate the usefulness of search sessions effectively.

\section{Discussion and Conclusion}
In this study, we explored the potential of LLMs in automated usefulness judgment tasks and assessed the performance of existing pre-trained models across different datasets and search contexts. Our findings indicate that while LLMs exhibit low to moderate performance in usefulness labeling, their effectiveness improves in shorter search sessions where they achieve higher correlation with human judgments. Additionally, larger models show a better grasp of contextual search relationships, consistently outperforming baseline approaches. Differing from traditional TREC-style relevance assessments, which primarily rely on query-document pair semantic understanding, usefulness judgment requires a more comprehensive understanding of user intents, tasks, and search contexts, making it inherently more complex and user-centered. A key limitation we encountered was the lack of a uniform dataset for usefulness evaluation, making it difficult to establish consistent benchmarks. Additionally, performance of pre-trained LLMs in usefulness labeling proved moderate, highlighting the need for more targeted fine-tuned adaptations in user-centered IR evaluation. This study opens a new research and application direction under \textit{LLMs-as-Judges} research~\cite{gu2024survey} by examining how LLMs can analyze search behavior logs and predict usefulness, and sheds light on both their potential and challenges. Built upon the findings, our future research will develop fine-tuning techniques and specialized models to enhance LLM performance in document usefulness labeling. Furthermore, we aim to implement models for automated usefulness judgment in text and conversational retrieval evaluation, real-time \textit{usefulness-based} re-ranking, and search path recommendations.
\begin{acks}
This work was partially supported by cloud computing credits provided through an Amazon Research Award to Dr. Chirag Shah.
\end{acks}

\bibliographystyle{ACM-Reference-Format}
\balance
\bibliography{reference, sample-base}

\end{document}